\def\lsi{\raise0.3ex\hbox{$<$\kern-0.75em\raise-1.1ex\hbox{$\sim$}}}
\def\gsi{\raise0.3ex\hbox{$>$\kern-0.75em\raise-1.1ex\hbox{$\sim$}}}

\documentclass{article}

 \usepackage{setspace}
\usepackage[sort&compress,square]{natbib}
\usepackage{graphicx}
\usepackage{geometry}
\title{The order of the quantum chromodynamics transition predicted 
by the standard model of particle physics}

\author{Y.~Aoki$^a$, G.~Endr\H{o}di$^b$, Z. Fodor$^{a,b}$, S.D.~Katz$^{a,b}$,
K.K. Szab\'o$^a$\\
$^a$Department of Physics, University of Wuppertal, D-42097 Wuppertal,
Germany.\\
$^b$Institute for Theoretical Physics, E\"otv\"os University,
H-1117 Budapest, Hungary.
}
\date{\today}

\begin{document}
\maketitle
%
{\bf 

Quantum chromodynamics (QCD) is the theory of the strong
interaction, explaining (for example) the binding of three almost
massless quarks into a much heavier proton or neutron -- and thus
most of the mass of the visible Universe. The standard model of
particle physics predicts a QCD-related transition that is relevant
for the evolution of the early Universe. At low temperatures, the
dominant degrees of freedom are colourless bound states of
hadrons (such as protons and pions). However, QCD is asymptotically
free, meaning that at high energies or temperatures the
interaction gets weaker and weaker~\cite{Gross:1973id,Politzer:1973fx}, 
causing hadrons to break
up. This behaviour underlies the predicted cosmological transition
between the low-temperature hadronic phase and a high-temperature
quark--gluon plasma phase (for simplicity, we use the
word 'phase' to characterize regions with different dominant
degrees of freedom). Despite enormous theoretical effort, the
nature of this finite-temperature QCD transition (that is, first-order,
second-order or analytic crossover) remains ambiguous.
Here we determine the nature of the QCD transition using
computationally demanding lattice calculations for physical
quark masses. Susceptibilities are extrapolated to vanishing lattice
spacing for three physical volumes, the smallest and largest of
which differ by a factor of five. This ensures that a true transition
should result in a dramatic increase of the susceptibilities.No such
behaviour is observed: our finite-size scaling analysis shows that
the finite-temperature QCD transition in the hot early Universe
was not a real phase transition, but an analytic crossover (involving
a rapid change, as opposed to a jump, as the temperature
varied). As such, it will be difficult to find experimental evidence
of this transition from astronomical observations.}

During the evolution of the Universe there were particle-physics-related 
transitions. Although there are strong indications of an inflationary 
period, we know little about how it affected possible 
transitions of our known physical model. To understand the 
consequences, we need a clear picture about these 
cosmologically relevant transitions. The Standard Model (SM) of particle 
physics predicts two such transitions.

One of the SM based transitions occurs at temperatures ($T$) of a few 
hundred GeV.  This transition is responsible for the spontaneous breaking 
of the electroweak symmetry, which gives the masses of the elementary 
particles. This transition is also related to the electroweak 
baryon-number violating processes, which had a major influence on the 
observed baryon-asymmetry of the Universe. Lattice results have shown 
that the electroweak transition in the SM is an analytic 
cross-over~\cite{Karsch:1996yh,Kajantie:1996mn,Csikor:1998eu,Gurtler:1997hr}.

The second transition occurs at $T\approx$200~MeV. It is 
related to the spontaneous breaking of the chiral symmetry of QCD. 
The nature of the QCD transition affects our understanding of the
Universe's evolution~(see Ref.~\cite{Schwarz:2003du} for example). 
In a strong 
first-order phase transition the quark--gluon plasma supercools before 
bubbles of hadron gas are formed. 
These bubbles grow, collide and merge, during which gravitational 
waves could be produced~\cite{Witten:1984rs}. Baryon-enriched 
nuggets could remain 
between the bubbles, contributing to dark matter.
The hadronic phase is the initial condition for nucleosynthesis, so 
inhomogeneities in this phase could have a strong 
effect on nucleosynthesis~\cite{Applegate:1985qt}. 
As the first-order phase transition weakens, these effects become less 
pronounced. Our calculations provide strong evidence that 
the QCD transition is a crossover and thus the 
above scenarios ---and many others---  are ruled out.
 
We emphasize that extensive 
experimental work is currently being done with heavy ion collisions 
to study the QCD transition (most recently at the Relativistic Heavy
Ion Collider, RHIC). Both for the cosmological transition
and for RHIC, the net baryon densities are quite small, and so the
baryonic chemical potentials ($\mu$) are much less than the typical 
hadron masses
($\approx$45~MeV at RHIC and negligible in the early Universe). 
A calculation at $\mu$=0 is directly applicable for the
cosmological transition and most probably also determines the 
nature of the transition at RHIC. Thus we carry out our analysis at
$\mu$=0. 

QCD is a generalised version of quantum electrodynamics (QED). The Euclidean 
Lagrangian with gauge coupling $g$ and with a quark mass of $m$ can be 
written as 
${\cal L}$=$-1/(2g^2)$Tr$F_{\mu\nu}$$F_{\mu\nu}$+${\bar \psi}$$\gamma_\mu$($\partial_\mu$+$A_\mu$+m)$\psi$, 
where
$F_{\mu\nu}$=$\partial_\mu$$A_\nu$-$\partial_\nu$$A_\mu$+[$A_\mu$,$A_\nu$]. 
In electrodynamics the gauge field $A_\mu$ is a simple real field, 
whereas in QCD it is a 
3$\times$3 matrix. Consequently  the commutator in $F_{\mu\nu}$ vanishes
for QED, but it does not vanish in QCD.  The $\psi$ fields also have 
an additional ``colour'' index in QCD, which runs from 1 to 3. 
Different types of quarks are represented by
fermionic fields with different masses. 
The action $S$ is defined as the four-volume integral of ${\cal L}$.
The basic quantity we determine is the the partition function
$Z$, which is the sum of the Boltzmann factors $\exp(-S)$ for all field
configurations. Partial derivatives of $Z$ with respect to the 
masses give rise to the order parameters we studied here.

There are some QCD results and model calculations to determine the 
order of the
transition at $\mu$=0 and $\mu$$\neq$0 for different fermionic contents 
(compare 
refs~\cite{Pisarski:1983ms,Celik:1983wz,Kogut:1982rt,Gottlieb:1985ug,Brown:1988qe,Fukugita:1989yb,Halasz:1998qr,Berges:1998rc,Schaefer:2004en,Herpay:2005yr}). 
Unfortunately, none of these
approaches can give an unambiguous answer for the order of the transition
for physical values of the quark masses. The only known systematic 
technique which could give a final
answer is lattice QCD.  

Lattice QCD discretises 
the above Lagrangian on a four-dimensional lattice 
and extrapolates the results
to vanishing lattice spacing ($a\longrightarrow0$).
A convenient way to carry out this discretisation is to
put the fermionic variables on the sites of the lattice, whereas
the gauge fields are treated as $3\times 3$ matrices connecting these 
sites. In this sense, lattice QCD is a classical four-dimensional
statistical physics system. One important difference compared to 
three dimensional systems is that the
temperature ($T$) is determined by the additional, Euclidean time extension 
($N_t$): $T$=1/($N_ta$).
Keeping the temperature fixed (such as at the
transition point) one can reduce $a$ and approach the 
continuum limit by increasing $N_t$
(see~{\bf Methods}).

There are several lattice results for the order of the QCD transition
(for the two most popular lattice fermion formulations see 
refs~\cite{Brown:1990ev} and~\cite{AliKhan:2000iz}), 
although they have unknown systematics.
We emphasise that from the lattice point of view 
two 'ingredients' are necessary to eliminate these systematic uncertainties.  

The first ingredient is to use physical values for the quark masses. 
Owing to the 
computational costs this is a great challenge in lattice QCD. Previous
analyses used computationally less demanding non-physically large quark
masses. However, these choices have limited relevance. The order of 
the transition depends on the quark mass. For example, in three-flavour QCD 
for vanishing quark masses the transition is of first-order. For 
intermediate masses it is most probably a crossover. For infinitely
heavy quark masses the transition is again first-order. 
For questions concerning the restoration
of chiral symmetry (such as the order of the transition), a controlled
extrapolation from larger quark masses (such as chiral
perturbation theory) is unavailable, and so the physical quark masses
should be used directly. 

The second ingredient is to remove the uncertainty associated with the 
lattice discretization.
Discretization errors disappear in the continuum limit; however,
they strongly influence the results at non-vanishing
lattice spacing. 
In three-flavour unimproved
staggered QCD, using a lattice spacing of about 0.28~fm, the first-order 
and the crossover regions are separated by a pseudoscalar mass of
$m_{\pi,c}$$\approx$300~MeV. Studying the same three-flavour theory
with the same lattice spacing, but with an improved 
p4 action  
(which has different discretization errors)  
we obtain $m_{\pi,c}$$\approx$70~MeV. In the first approximation,
a pseudoscalar mass of 140~MeV (which corresponds to the numerical
value of the physical pion mass) 
would be in the first-order transition region, whereas using the
second approximation, it would be in the  crossover region.
The different discretisation uncertainties are solely responsible for 
these qualitatively different results~\cite{Karsch:2003va}. 
Therefore, the proper approach is to use physical quark masses,  
and to extrapolate to vanishing lattice spacings. 
Our work eliminates both of the above uncertainties.

Our goal is to identify the nature of the transition 
for physical quark masses as
we approach the continuum limit. 
We will study the finite size scaling of the 
lattice chiral susceptibilities 
$\chi(N_s,N_t)$=$\partial^2$/$(\partial$$m_{ud}^2)$($T/V$)$\cdot\log Z$,
where $m_{ud}$ is the mass of the light u,d quarks and $N_s$ is the spatial
extension.
This susceptibility shows a pronounced
peak around the transition temperature ($T_c$). For a real phase transition
the height of the susceptibility peak increases and
the width of the peak decreases when we increase the volume.
For a first-order phase transition the finite size scaling is determined by
the geometric dimension, the height is proportional to $V$, and the
width is proportional to $1/V$. For a second-order transition the 
singular behaviour is given by some power of $V$, defined by the 
critical exponents. The picture would be completely different for an
analytic crossover. There would be no singular behaviour and the
susceptibility peak does not get sharper when we increase the volume;
instead, its height and width will be $V$ independent for large volumes.

\begin{figure}[h!]
\centerline{\includegraphics*[bb=20 430 580 710,width=13cm]{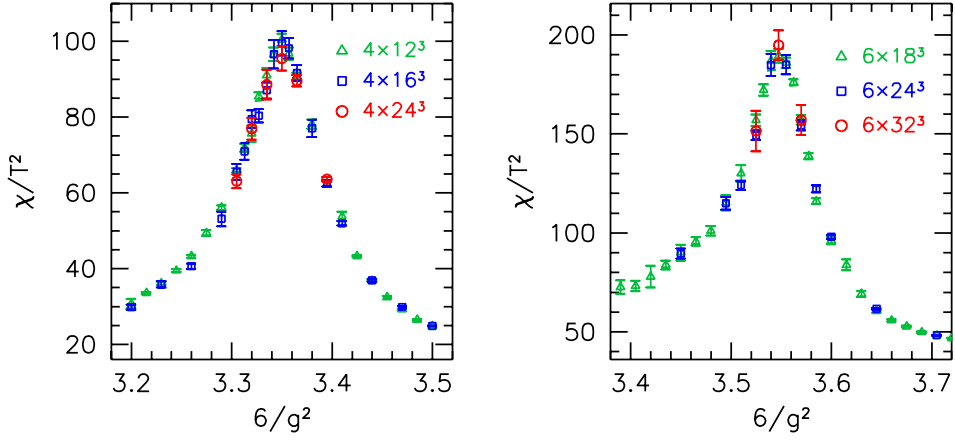}}
\caption{\label{susc_46}
Susceptibilities  for the light quarks for $N_t$=4 (left panel) and
for $N_t$=6 (right panel) as a function of $6/g^2$, where $g$ is the gauge coupling
($T$ grows with $6/g^2$).
The largest volume
is eight times bigger than the smallest one, so a first-order phase
transition would predict a susceptibility peak that is eight times
higher (for
a second-order phase transition the increase would be somewhat less,
but still dramatic). Instead of such a significant change we
do not observe any volume dependence. Error bars are s.e.m.
}
\end{figure}

Figure~\ref{susc_46} shows the susceptibilities for the light quarks 
for $N_t=$4 and 6, for which we used aspect ratios $r=N_s/N_t$ 
ranging from 3 to 6 and 3 to 5, respectively. 
A clear signal for an analytic 
crossover for both lattice spacings can be seen. However,
these curves do not say much about the continuum behaviour
of the theory. In principle a phenomenon as unfortunate as that in
the three-flavour theory could occur~\cite{Karsch:2003va}, in which the 
reduction of the 
discretization effects changed the nature of the transition for
a pseudoscalar mass of $\approx$140~MeV.

Because we are interested in genuine
temperature effects we subtract the $T$=0 susceptibility
and study only the difference between $T$$\neq$0 and $T$=0 at 
different lattice spacings.
To do it properly, when we approach the continuum 
limit the renormalization of $\chi$ has to be performed. This
leads to $m^2$$\Delta$$\chi$, which we study (see {\bf Methods}). 

\begin{figure}[h!]
\centerline{\includegraphics*[bb=20 500 592 690,width=17cm]{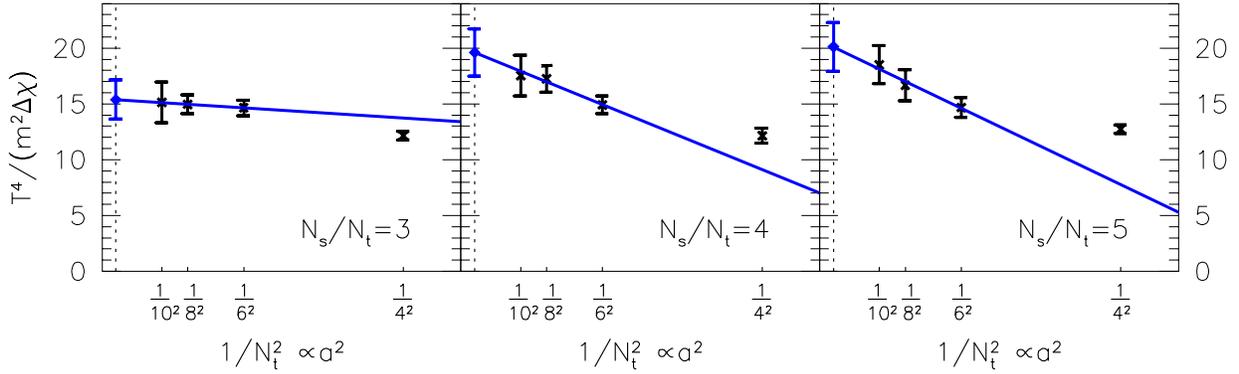}}
\caption{\label{cont_ex}
Normalised susceptibilities $T^4/(m^2\Delta\chi)$ 
for the light quarks for aspect
ratios r=3 (left panel) r=4 (middle panel) and r=5 (right panel)
as functions of the lattice spacing. Continuum 
extrapolations are carried out for all three physical volumes and the
results are given by the leftmost blue diamonds. Error bars are s.e.m with 
systematic estimates.
}
\end{figure}

To give a continuum result for the order of the transition we
carry out a finite size scaling analysis of the dimensionless quantity
$T^4/(m^2\Delta\chi)$ directly in the
continuum limit. For this study we need the height of the susceptibility
peaks in the continuum limit for fixed physical volumes.
The continuum extrapolations are done
using four different lattice spacings ($N_t$=4,6,8 and 10). The volumes
at different lattice spacings are fixed in units of $T_c$, and thus
$VT_c^3$=$3^3$,$4^3$ and $5^3$ were chosen.
(In three cases the computer architecture did
not allow us to take the above ratios directly. In these
cases, we used the next possible volume and interpolated or
extrapolated. The height of the peak depends weakly
on the volume, so these procedures were always safe.)
Altogether we used twelve different lattice volumes ranging from 
$4\cdot12^3$ to $10\cdot48^3$ at $T>0$. 
For the $T=0$ runs lattice volumes 
from $24\cdot12^3$ up to $56\cdot28^3$ were used.
The number of trajectories were between 1500 and 8000 for $T>0$ 
and between 1500 and 3000 for $T=0$, respectively.
Figure~\ref{cont_ex} shows the continuum extrapolation for
the three different physical volumes. The $N_t$=4 results are
slightly off but the $N_t$=6,8 and 10 results show a good
$a^2$$\propto$$1/N_t^2$ scaling.

\begin{figure}[h!]
\centerline{\includegraphics*[width=10cm,bb=0 170 592 614]{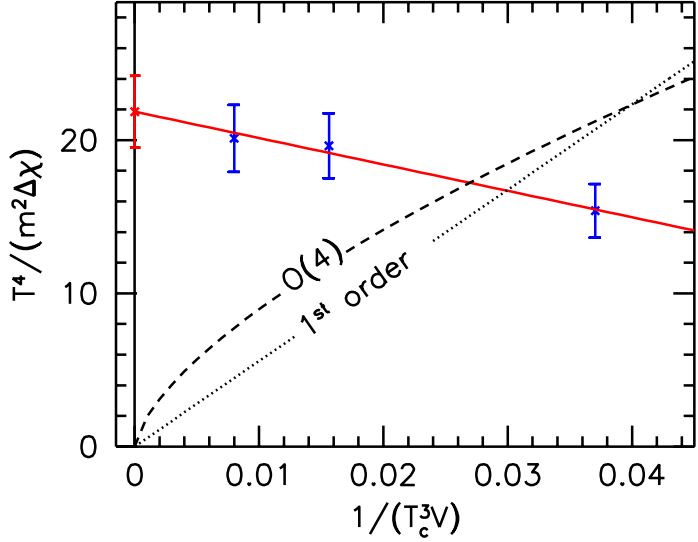}}
\caption{\label{cont_scal}
Continuum extrapolated susceptibilities $T^4/(m^2\Delta\chi)$ 
as a function of 1/$(T_c^3V)$.
For true phase transitions the infinite volume extrapolation should
be consistent with zero, whereas
for an analytic crossover the infinite volume extrapolation gives
a non-vanishing value. The continuum-extrapolated
susceptibilities show no phase-transition-like volume dependence, though
the volume changes by a factor of five.
The V$\rightarrow$$\infty$ extrapolated value is 22(2)
which is
11$\sigma$ away from zero. For illustration, we fit the expected 
asymptotic
behaviour for first-order and O(4) (second order) 
phase transitions shown by dotted and dashed lines,
which results in chance probabilities of 
$10^{-19}$ ($7\times10^{-13}$), respectively. Error bars are s.e.m with 
systematic estimates.
}
\end{figure}

Having obtained the continuum values for 
$T^4/(m^2\Delta\chi)$ at fixed physical volumes, we study the
finite size scaling of the results. Figure~\ref{cont_scal} shows our 
final results.
The volume dependence strongly suggests that there is no true phase 
transition but only an analytic crossover in QCD.

{\bf Methods.} An introduction to lattice QCD at $T$=0 and $T$$\neq$0
can be found in refs~\cite{Davies:2002cx} and~\cite{Ukawa:1995tc} for example.  
The detailed form of our Symanzik improved gauge and stout-link improved 
staggered fermionic action can be found in ref.~\cite{Aoki:2005vt}.
(Staggered QCD with $n_f$$\neq$4
flavours uses the fourth-root description for the fermionic determinant,
the relevance of which is recently intensively discussed, see 
ref.~\cite{Bernard:2006ee} and references therein.) 
In this work we also used a stout-smearing level of 2.  
Note that stout-link improvement makes the staggered fermion 
taste symmetry violation small even at moderate lattice spacings 
(for an illustration, see Fig. 1 of ref.~\cite{Aoki:2005vt}). 
Taste symmetry violation errors --as with other discretization errors of 
the staggered formalism-- scale as $a^2$ and disappear only after 
extrapolating to the continuum limit. We carried out this extrapolation 
and explicitly checked that taste symmetry violations for our
smallest lattice spacings are 
already within this $a^2$ scaling regime, and so we conclude 
that the extrapolation is reliable.

In previous staggered analyses the gauge configurations were produced
by the R-algorithm~\cite{Gottlieb:1987mq}
at a given step size. The step size is an
intrinsic parameter of the algorithm, which has to be extrapolated to zero.
Instead of using
the approximate R-algorithm, we used~\cite{Aoki:2005vt}
the exact rational hybrid Monte-Carlo 
algorithm~\cite{Clark:2003na} in large scale
simulations.
This technique, which we apply also in this paper, expresses
the fractional powers of the Dirac operator by rational functions.

\begin{figure}[h!]
\centerline{\includegraphics*[width=10cm]{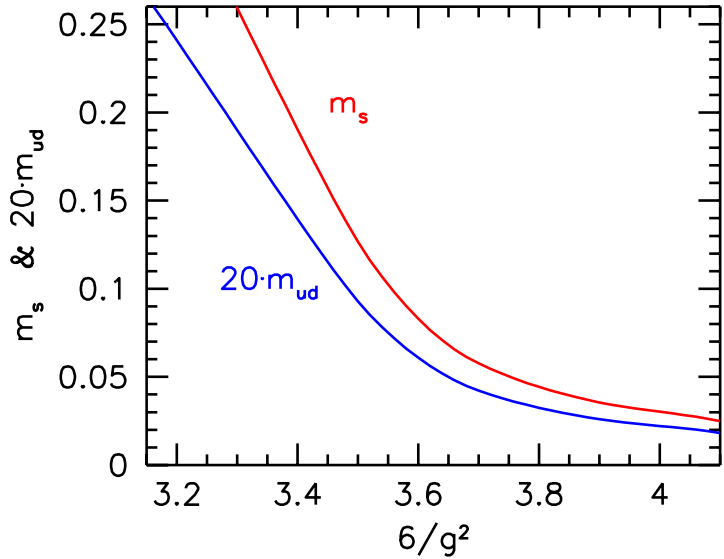}}
\caption{\label{lcp}
The line of constant physics. 
We show our choice for $m_s$  (strange quark mass) and 20$m_{ud}$
(u,d quark masses) in lattice units as functions of $6/g^2$. 
}
\end{figure}

In our simulations we approach the continuum limit along the line
of constant physics (LCP). The LCP is defined by relationships between 
the bare lattice parameters (gauge coupling $g$ and lattice bare 
quark masses for light quarks $m_{ud}$
and strange quark $m_s$). These relationships ensure that the physics 
(such as ratios of physical observables) remain constant, while 
changing any of
the parameters. We note that the LCP is unambiguous
(independent of the physical quantities, which are used to define the above
relationships) only in the vicinity of the 
continuum limit.
The present work uses the LCP
defined by fixing two ratios to their physical values:
$m_K$/$m_\pi$=3.689 and $f_K$/$m_\pi$=1.185 
($m_K$ is the mass of the kaon, $f_K$ is its decay constant and 
$m_\pi$ is the mass of the pion). Figure \ref{lcp} shows the LCP used
in this work.

Using the above action, simulation algorithm and parameter set 
based on our LCP, we performed simulations at  $T$$\neq$0 and $T$=0.
At $T$$\neq$0 we always used the physical quark masses
at four different temporal extensions: $N_t$=4,6,8 and 10 lattices. 
The aspect ratios ($N_s$/$N_t$) were taken to be 3,4 and 5 (for 
the $N_t=4$ case also 6), which
results in a factor-of-8  change in the volume ($V$). 
In the chirally broken phase (our zero temperature
simulations always belong to
this class) chiral perturbation theory can be used to
extrapolate 
to the physical values of the light quark masses
in a controlled manner. 
Therefore we used four
pion masses ($m_\pi\approx$235, 300, 355 and 405~MeV), 
which are somewhat larger than
the physical one. The spatial extension was chosen
to ensure $N_s$$m_\pi$$\gsi$4. The computational requirement for 
the present work was 0.8 teraflopyears.

The renormalization of the chiral susceptibility 
can be done by taking the second derivative of  
the free energy density ($f$) with respect to the renormalized mass 
($m_r$). 
We apply the usual definition:
$f/T^4=-N_t^4[\log Z(N_s,N_t)/(N_t N_s^3)-\log Z(N_{s0},N_{t0})/(N_{t0} N_{s0}^3)]$. 
This quantity has a correct 
continuum limit. The subtraction term is obtained at $T$=0, for which  
simulations are carried out on lattices with $N_{s0}$, $N_{t0}$ spatial 
and temporal extensions (otherwise at the same parameters of the action). 
The bare light quark mass is related to $m_r$ by the
mass renormalisation constant $m_r$=$Z_m$$\cdot$$m_{ud}$.  
Note that $Z_m$
falls out of the combination 
$m_r^2$$\partial^2$/$\partial$$m_r^2$=$m_{ud}^2$$\partial^2$/$\partial$$m_{ud}^2$.
Thus, $m_{ud}^2\left[\chi(N_s,N_t)-\chi(N_{s0},N_{t0})\right]$ also has a 
continuum limit 
(for its maximum values for different $N_t$, and in the continuum limit
we use the shorthand 
notation $m^2$$\Delta \chi$).
For the subtraction, $T$=0 simulations are performed and chiral perturbation 
theory is applied.
 
\bibliographystyle{unsrt}
\bibliography{pt_9}

\vspace*{2cm}

{\bf Acknowledgements}
We thank F.~Csikor, A.~Dougall, K.-H. Kampert, M. Nagy, Z.~R\'acz and 
D.J. Schwarz
for discussions.
This research was partially supported by a DFG German Science Grant, 
OTKA Hungarian Science Grants and an EU research grant.
The computations were carried out 
on PC clusters at the University of Budapest and Wuppertal with
next-neighbour communication architecture~\cite{Fodor:2002zi} and on the 
BlueGene/L machine in J\"ulich.
A modified version of the publicly available MILC code
(http://physics.indiana.edu/$\tilde{\ }$sg/milc.html) was used.

\newpage

\noindent{\bf Figures}

\vspace*{2cm}


\end{document}